\pdfoutput=1

\documentclass[a4paper]{article}
\RequirePackage{etex}

\usepackage[left=1.5cm,right=1.5cm,top=1.5cm,bottom=1.5cm,ignoreheadfoot]{geometry}

\usepackage[utf8]{inputenc}
\usepackage{pgfplots}
\pgfplotsset{compat=newest}
\usepackage[title]{appendix}
\usepackage{amssymb}
\usepackage{booktabs}
\usepackage{textcomp}
\usepackage{graphicx}
\usepackage{physics}
\usepackage{mathtools}
\usepackage[export]{adjustbox}
\usepackage{yhmath}
\usepackage{makecell}
\usepackage{amsmath}
\usepackage{wrapfig}
\usepackage{xparse}
\usepackage{xargs,ifthen,xstring,etoolbox}
\usepackage{environ}
\usepackage{multirow}
\usepackage{array}
\usepackage{hyperref}
\usepackage{amsfonts}
\usepackage{dsfont}
\usepackage{bbm}
\usepackage{caption}
\usepackage{subcaption}
\usepackage{xspace}
\usepackage{xcolor,colortbl}
\usepackage{tcolorbox}
\usepackage{authblk}
\usepackage{soul}
\usepackage{csquotes}
\usepackage{tikz-network}

\usepackage{doi}
\usepackage[numbers]{natbib}
\newcommand\Order[1]{\ensuremath{\mathcal{O}\left(#1\right)}}

\usepackage{tikz}

\usetikzlibrary{patterns,calc,matrix,backgrounds,fadings,shapes,arrows,shadows,decorations.pathreplacing,shadows.blur}
\usetikzlibrary{trees}
\usetikzlibrary{quantikz2}
\usetikzlibrary{tikzmark}

\usepgfplotslibrary{external} 
\usepackage{amsthm}
\usepackage{letltxmacro}
\LetLtxMacro\amsproof\proof
\LetLtxMacro\amsendproof\endproof
\usepackage{thmbox}
\AtBeginDocument{%
  \LetLtxMacro\proof\amsproof
  \LetLtxMacro\endproof\amsendproof
}

\newcommand{\clr}[0]{\operatorname{clr}}

\newcommand\R[0]{\ensuremath{\mathbb{R}}}
\newcommand\N[0]{\ensuremath{\mathbb{N}}}

\renewcommand\b[0]{\mathbf{b}}
\renewcommand\c[0]{\mathbf{c}}
\renewcommand\t[0]{\mathbf{t}}

\newcommand{\overbar}[1]{\mkern 1.5mu\overline{\mkern-1.5mu#1\mkern-1.5mu}\mkern 1.5mu}
\newcommand{\lX}[0]{\overbar{X}}
\newcommand{\maxkcut}[1]{MAX $#1$-CUT}
\newcommand{\maxcut}{MAX ($2$-)CUT\xspace}
\newcommand{\Span}[1]{\operatorname{span}\!\left(#1\right)}
\newcommand\restr[2]{{
    \left.\kern-\nulldelimiterspace 
    #1 
    \vphantom{\big|} 
    \right|_{#2} 
}}

\newcommand{\datfilename}[7]{tables/#1_k#2_direct#3_forceP2#4_fix_one_node#5_color_encoding#6_problem_encodingbinary_shots100000_mixer#7.dat}

\newcommand{\imshow}[2]{
    \adjustbox{valign=m,raise=-.25cm}{
    \begin{tikzpicture}
        \begin{axis}[
        scale=#2,
            axis on top=true,
            x label style={at={(axis description cs:1.,0.025)},anchor=north},
            y label style={at={(axis description cs:-0.2,0.85)},rotate=-90,anchor=south},
            scale only axis,
            xlabel=$\gamma/\pi$,
            ylabel=$\beta/\pi$,
            ymin=-0.05,
            ymax=1.95,
            xmin=-0.05,
            xmax=1.95,
            xtick = {0,1,2},
            ytick = {0,1,2},
            unit vector ratio={1 1},
            colormap/hot2,
            colorbar,
            colorbar style={
                tick label style={/pgf/number format/fixed}, 
                at={(1.05,1.0)},
                width=0.05*\pgfkeysvalueof{/pgfplots/parent axis width},
                ytick distance=0.1, 
                samples=3,         
            },
            ]
            \addplot [
                matrix plot*,
                point meta=explicit
            ] table [meta index=2] {#1};
        \end{axis}
    \end{tikzpicture}
    }
}

\definecolor{verylightgray}{gray}{0.95}

\title{Encodings of the weighted \maxkcut{k} on qubit systems}

\author[$\dagger$]{Franz G. Fuchs}
\author[$\dagger$]{Ruben P. Bassa}
\author[$\ddagger$]{Frida Lien}

\affil[$\dagger$]{SINTEF AS, Department of Mathematics and Cybernetics, Oslo, Norway}
\affil[$\ddagger$]{University of Oslo, Department of Mathematics, Oslo, Norway}

\date{\today}

\begin{document}

\maketitle
\begin{abstract}
The weighted \maxkcut{k} problem involves partitioning a weighted undirected graph into k subsets, or colors, to maximize the sum of the weights of edges between vertices in different subsets.
This problem has significant applications across multiple domains.
This paper explores encoding methods for \maxkcut{k} on qubit systems, utilizing quantum approximate optimization algorithms (QAOA) and addressing the challenge of encoding integer values on quantum devices with binary variables.
We examine various encoding schemes and evaluate the efficiency of these approaches.
The paper presents a systematic and resource efficient method to implement phase separation for diagonal square binary matrices.
When encoding the problem into the full Hilbert space, we show the importance of encoding the colors in a balanced way.
We also explore the option to encode the problem into a suitable subspace, by designing suitable state preparations and constrained mixers (LX- and Grover-mixer).
Numerical simulations on weighted and unweighted graph instances demonstrate the effectiveness of these encoding schemes, particularly in optimizing circuit depth, approximation ratios, and computational efficiency.
\end{abstract}

\section{Introduction and related work}

The quantum approximate optimization algorithm~\cite{farhi2014quantum}/quantum alternating operator ansatz~\cite{hadfield2019quantum} (QAOA) is a widely studied hybrid algorithm for approximately solving combinatorial optimization problems encoded into the ground state of a Hamiltonian.
Given an objective function $f:\{0,1\}^n \rightarrow \R$, the problem Hamiltonian is defined as $H_P\ket{x} = f(x)\ket{x}$. Thus, the ground states of $H_P$ correspond to the minima of the objective function.
In the general framework of QAOA, a parameterized quantum state is prepared as follows:
\begin{equation*}
    \ket{\gamma,\beta} = U_M(\beta_p) U_P(\gamma_p) \cdots U_M(\beta_1) U_P(\gamma_1) \ket{\phi_0}.
\end{equation*}
Starting from an initial state $\ket{\phi_0}$, the algorithm alternates between applying the phase-separation operator $U_P(\gamma)$ and the mixing operator $U_M(\beta)$. The goal is to optimize the parameters $\gamma, \beta \in \R$ such that $\bra{\gamma,\beta} H_P \ket{\gamma,\beta}$ is minimized.

Since the original proposal of QAOA, several extensions and variants have been introduced. One such extension is ADAPT-QAOA~\cite{zhu2022adaptive}, an iterative version that iteratively adds only the terms that contribute to lowering the expectation value the most.
Another variant is R-QAOA~\cite{bravyi2019obstacles}, which progressively removes variables from the Hamiltonian until the reduced instance can be classically solvable.
A third extension, WS-QAOA~\cite{egger2020warm}, incorporates solutions from classical algorithms to warm-start the QAOA process.
When a given problem has constraints, the design of mixers 
is vital.
For constrained problems, the design of mixers is crucial. A well-known example is the $XY$-mixer~\cite{hadfield2017quantum, Wang2020}, which restricts evolution to "k-hot" states. GM-QAOA~\cite{Bartschi_2020} uses Grover-like operators for all-to-all mixing of more general feasible subsets. 
The mixing operator of GM is comprised of a circuit $U_S$ that prepares the uniform superposition of all feasible states, its inverse $U_S^\dagger$
and a multi-controlled rotational phase shift gate.
Inspired by the stabilizer formalism, the LX-Mixer~\cite{fuchs2023LX,Fuchs_2022} is a generalization of the aforementioned methods that provides a flexible framework 
that allows for tailored mixing within the feasible subspace.

To increase the practical relevance of QAOA, encoding problems with integer values, rather than binary variables, into qubit systems is an important research area.
This process generally introduces considerable flexibility in en-
\begin{wrapfigure}{l}{0.25\textwidth}
    \centering
    \vspace{-1\baselineskip}
    \includegraphics[width=0.8\linewidth]{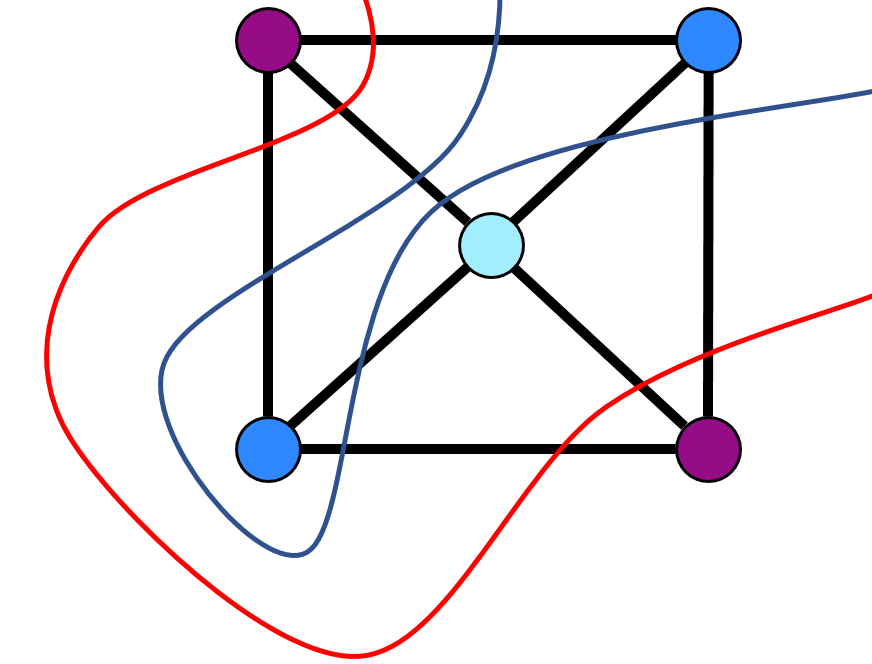}
    \caption{An example of an optimal solution for a \maxkcut{3} problem.}
    \label{fig:example_max4cut}
    \vspace{-1\baselineskip}
\end{wrapfigure}
coding choices, affecting not only the number of qubits and controlled operations needed, but also the complexity 
of the optimization landscape and the quality of the approximation ratios that can be achieved.
In this paper, we explore various encoding schemes for the \maxkcut{k} problem into qubit systems, building on and improving earlier work~\cite{fuchs2021efficient}.
This problem has numerous applications, including the placement of television commercials
in program breaks, arrangement of containers on a ship with $k$ bays, partition a set of items (e.g. books sold by an online shop) into $k$ subsets, product module design, frequency assignment, scheduling, and pattern matching~\cite{Aardal2007,gaur2008capacitated}.
Given a weighted undirected graph $G=(V,E)$, the \maxkcut{k} problem consists of finding a maximum-weight $k$-cut.
This involves partitioning the vertices into $k$ subsets such that the sum of the weights of the edges between vertices in different subsets is maximized, as illustrated in Figure~\ref{fig:example_max4cut}.
By assigning a label $x_i\in \{1, \dots, k\}$ to each vertex $i\in V$ and
defining $\mathbf{x} = (x_1,\ldots,x_{|V|})$, the optimization problem for \maxkcut{k} can be formulated as
\begin{equation}
\label{eq:maxkcolor}
    \underset{\mathbf{x}\in\{1,\dots,k\}^n}{\max} C(\mathbf{x}), \qquad  C(\mathbf{x}) = \sum_{(i,j)\in E} 
    \begin{cases}
        w_{ij}, & \text{ if } x_i \neq x_j\\
        0, & \text{otherwise}.
    \end{cases}
\end{equation}
Here, $w_{ij}>0$ is the weight of edge $(i,j)\in E$.
The function $C(\mathbf{x})$ is commonly referred to as the cost function.
For $k=2$ the \maxkcut{k} problem becomes the well known MAX CUT problem.
The \maxkcut{k} problem is NP-complete,
and it has been demonstrated that no polynomial-time approximation scheme exists for $k\geq2$, unless P=NP \cite{frieze1997improved}.
A randomized approximation algorithm for \eqref{eq:maxkcolor} has an approximation ratio $\alpha$ if
\begin{equation*}
    \mathbb{E}[C(\mathbf{x})] \geq \alpha C(\mathbf{x}^*),
\end{equation*}
where $\mathbf{x}^*$ denotes an optimal solution. 

\begin{figure}
    \centering
    \begin{tikzpicture}[
edge from parent fork down
]
    \tikzset{level 1/.style={sibling distance=7cm, level distance = 1cm}},
    \node {\maxkcut{k} qubit encodings}
    child [level 2/.style={sibling distance=5.5cm, level distance = 1cm},
           ] {
        node[yshift=.0cm] {one-hot}
        child [level 3/.style={sibling distance=5.5cm, level distance = 2cm},
           ]{
           child{
            node [draw, rounded corners, inner sep=3pt, text width=3.0cm] {
            \vspace{-1\baselineskip}
            \centering
            \phantom{\underline{Section:}}
            \vspace{.2\baselineskip}
            \begin{align*}
                    V&
                    \subsetneqq \mathcal{H}\\
                    H_M&= \text{LX}\\
                    H_P&=H_P^\text{one-hot}(k)
            \end{align*}
            \vspace{-1.8\baselineskip}
                    }
                }
                }
    }
    child [level 2/.style={sibling distance=6.0cm, level distance = 1cm},
           ] {
        node[yshift=.0cm]  {binary}
        child  [level 3/.style={sibling distance=4.0cm, level distance=2cm},
           ]{
        node[inner sep=3pt, yshift=0cm, fill=white] {$k=2^L$}
        child { node [draw, rounded corners, inner sep=3pt,text width=3.0cm] {
            \vspace{-1\baselineskip}
            \centering
            \underline{Section~\ref{sec:poweroftwo}:}
            \vspace{.2\baselineskip}
            \begin{align*}
                    V&=\mathcal{H}\\
                    H_M&= \text{X}\\
                    H_P&=H_P^\text{bin}(k)
            \end{align*}
            \vspace{-1.8\baselineskip}
        }
        }
        }
        child [level 3/.style={sibling distance=4.0cm, level distance=2cm},
           ]{
        node[inner sep=3pt, yshift=0cm, fill=white] {$k\neq2^L$}
            child
            [
         edge from parent path={(\tikzparentnode.south) -- +(0,-9pt) -| (\tikzchildnode.north)}
         ]
            { node [draw, rounded corners, inner sep=3pt,text width=3.0cm] {
                \vspace{-1\baselineskip}
                \centering
                \underline{Section~\ref{sec:notp2fullH}:}
                \vspace{.2\baselineskip}
                \begin{align*}
                        V&=\mathcal{H}\\
                        H_M&= \text{X}\\
                        H_P&=H_P^\text{bin}(k)
                \end{align*}
                \vspace{-1.8\baselineskip}
            }
            }
            child
            [
         edge from parent path={(\tikzparentnode.south) -- +(0,-9pt) -| (\tikzchildnode.north)}
         ]
            { node [draw, rounded corners, inner sep=3pt,text width=3.0cm] {
                \vspace{-1\baselineskip}
                \centering
                \underline{Section~\ref{sec:notp2subspace}:}
                \vspace{.2\baselineskip}
                \begin{align*}
                        V&
                        \subsetneqq \mathcal{H}\\
                        H_M&= \text{LX}\\
                        H_P&=H_P^\text{bin}(2^{L_k})
                \end{align*}
                \vspace{-1.8\baselineskip}
            }
            }
        }
    }
;
\end{tikzpicture}
    \caption{Overview of different ways to encode the problem.
    }
    \label{fig:overview_encodings}
\end{figure}
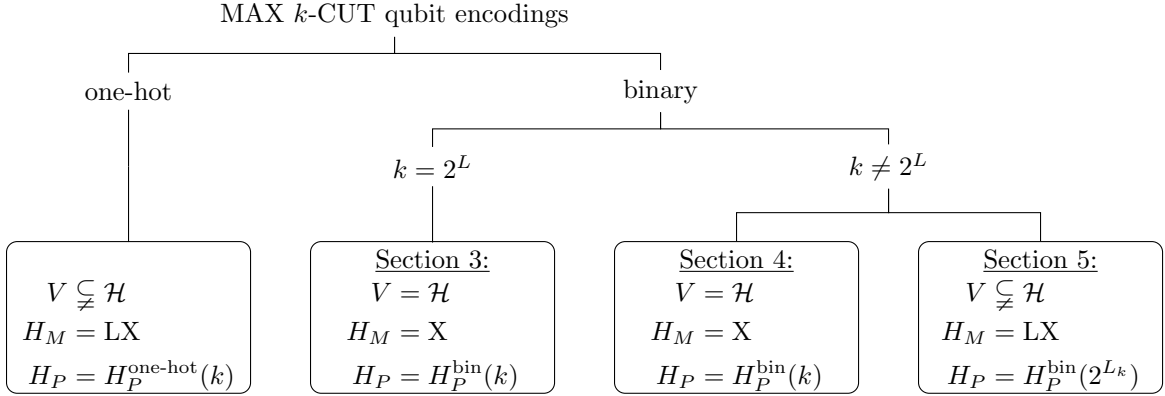

Expressing the solutions as strings of $k$-dits as in Equation~\eqref{eq:maxkcolor} is a natural extension of the \maxcut problem to $k>2$.
However, most quantum devices work with a two-level system, so we need to make a suitable encoding on qubits system~\cite{fuchs2021efficient,hadfield2019quantum}.
\emph{One-hot encoding} uses $k$ bits for each vertex, where the single bit that is $1$ encodes which set/color the vertex belongs to.
Using this encoding requires $k|V|$ qubits.
However, when $k$ is not a power of two, this results in more computational states per vertex than there are colors.
The corresponding subspaces need to be dealt with in a suitable fashion.
The Grover or the LX-mixer (which becomes the XY-mixer in this case) can be readily applied.
However, the ratio of the dimension of the feasible states to the system size is $k/2^k$ which becomes exponentially small for increasing $k$.
Therefore, it is advantageous to use a \emph{binary encoding}, which does not have this problem. For a given $k$ we encode the information of a vertex belonging to one of the sets by $\ket{i}_{n_k}$, which requires
\begin{equation*}
    n_k \coloneq \lceil log_2(k)\rceil
\end{equation*}
qubits.
Here $\lceil \cdot \rceil$ means rounding up to the nearest integer.
Binary encoding requires $n_k |V|$ qubits, which is exponentially fewer than required by one-hot encoding.
The resulting problem Hamiltonian can be written as the sum of local terms, i.e.
\begin{equation*}
    H_P = \sum_{e\in E} w_{e} H_{e},
\end{equation*}
where $w_{i,j}$ is the weight of the edge between vertices $i$ and $j$.
Furthermore, the local term can be expressed as
\begin{equation}
\label{eq:Hij}
    H_{e} = 2 \widehat H_e - I, \ \widehat H_e
     = \sum_{(i,j) \in \clr } \ket{i}\bra{i} \otimes \ket{j}\bra{j} .
\end{equation}
where $\clr$ is a set of index pairs $(i,j)$ representing \emph{equivalent colors} under an equivalence relation.
From the definition $H_e$ has eigenvalue $+1$ for a state $\ket{i}\ket{j}$ if $i \sim_{\clr} j $ and $-1$ if not.
The phase separating operator $e^{-it H_e}$ differs from $e^{-it \widehat H_e}$ only by a scalar factor and a global phase, so it is sufficient to realize the latter.

More formally, an equivalence relation $\sim_{\clr}$ is defined on $X = \{0, \ldots, 2^{n_k} - 1\}$, where $\clr$ consists of all pairs $(i,j)$ such that $i \sim_{\clr} j$.
For example, if
$
\clr = \{(0,0), (1,1), (2,2), (3,3), (2,3), (3,2)\},
$
the equivalence relation defines the following equivalence classes:
$
[0] = \{0\}, [1] = \{1\}, [2] = \{2, 3\}.
$
For convenience, we can work with a generating set.
Suppose $R$ is a relation on the set $X$.
Then the equivalence relation on $X$ generated by a $R$, denoted $\overbar{R}$, is the smallest equivalence relation on $X$ that contains $R$,
which can be obtained by the reflexive, symmetric, and transitive closure (in that order).
Hence, only the non-trivial relations must be specified, and the above example can be generated by the closure of the relation
$R=\{(2,3)\}$ which generates $\overbar{R} = \{(0,0), (1,1), (2,2), (3,3), (2,3), (3,2)\}$.
The \emph{set of colors} in the context of \maxkcut{k} is then the set of equivalence classes induced by $\sim_{\clr}$, denoted
\[
X / {\sim_{\clr}} := \{ [x] : x \in X \},
\]
with each equivalence class given by
\[
[c] := \{ x \in X : c \sim_{\clr} x \}.
\]

We recognize that the phase separating operator, i.e. the real time evolution of $\widehat H_e$ acts by applying a phase on a specific subsets of all computational basis states. When these subsets are sparse, such unitaries can be efficiently realized~\cite{Fuchs:2025eqs}:
Let $P_B$ be the projector onto a subspace given by set of computational basis states $B$.
The unitary operator $e^{itP_B}$ can be realized by a circuit for mapping 
the (indexed) states of $B$ to the first $|B|$ computational basis states
which can be achieved by
$
    M=\prod_{j=0}^{|B|-1} T_{j,z_j},
$
where $T_{i,j}$ is a transposition gate.
It follows directly that
\begin{equation*}
    M P_B M^\dagger \ket{j} = M P_B \ket{z_j} = 
    \begin{cases}
        M \ket{z_j} = \ket{j},& \text{ for } j< |B|,\\
        \mathbf{0},& \text{ otherwise.}
    \end{cases}
\end{equation*}
Therefore, we have that
$
    MP_{B}M^{\dag} 
    = \sum_{j< |B|}\ket{j}\bra{j},
$
from which it follows that the exponential has the form
\begin{equation*}
    e^{itP_B} =
    M^\dag e^{it\sum_{j< |B|}\ket{j}\bra{j}} M =
    M^\dag C_{\leq |B|} Ph(t) M,
\end{equation*}
where $C_{\leq |B|} Ph(t)$ is a low-pass controlled phase shift gate as defined in~\cite{Fuchs:2025eqs}.
For a general subspace, the overall circuit requires $\Order{n |B|}$ CX gates and $\Order{n |B| + \log(|B|) \log (\frac{1}{\epsilon})}$ T gates.

When the subspace is given by the span of a Pauli X-orbit as defined in~\cite{Fuchs:2025eqs}, the overall circuit requires only  $\Order{n \log(|B|)}$ CX gates and  $\Order{n+\log(\frac{1}{\epsilon})}$ T gates.
Here, $\epsilon$ is a given tolerance for approximating rotational gates.
Even though $\Span{B}$ might not be a Pauli X-orbit subspace, we can always decompose into a direct sum of Pauli X-orbit subspaces and apply the construction individually on the subspaces of the decomposition. 
We will therefor in this paper aim to divide the subspaces such that they are Pauli X-orbit subspaces if it leads to an efficient circuit.

An overview of possible encoding techniques is provided in Figure~\ref{fig:overview_encodings}.
When $k$ is not a power of two, we have to deal with the fact that we have non-trivial equivalence classes $\clr$.
There are two ways to deal with this. The first is to directly implement the phase separating Hamiltonian on the full Hilbert space~\cite{fuchs2021efficient}, or to restrict the evolution onto the relevant subspace.
The main contributions of this article are for the cases when $k$ is not a power of two.
\begin{itemize}
    \item We introduce a systematic way to implement the phase separating Hamiltonian, resulting in a reduction of the complexity of the resulting circuit using the theorems from \cite{Fuchs:2025eqs}.
    \item A method to directly work with a feasible subspace with a suitable state preparation and constrained mixer, see Section~\ref{sec:notp2subspace}.
    \item More balanced sets of equivalent colors, which help to improve the resulting cost landscape, see Section~\ref{sec:notp2fullH}.
\end{itemize}
We also employ the Grover mixer to all cases, indicating that the resulting landscape might be favourable.
Throughout the paper the notation $C^l U$ means an $l-$controlled $U$ gate. E.g. a Toffoli gate is $C^2X$ and $CPh=CPh(t)$ is the controlled phase shift gate, where we mostly omit the parameter.
We will make use of the following notation for \emph{multi-controlled unitary (MCU)} gates
\begin{equation}
    \label{eq:MCU}
    C^\c_\b U_\t =
    (I_\c - \ket{\b}\bra{\b}_\c) \otimes I_\t
    + \ket{\b}\bra{\b}_\c \otimes U_\t,
\end{equation}
where $\mathbf{c}, \mathbf{t}$ index control and target qubits, and $\mathbf{b}$ specifies the control condition.
    
\section{Binary encodings when k is a power of two}\label{sec:poweroftwo}

\begin{table}
    \centering
        \begin{tabular}{llrrr}
        \toprule
        & mixer & $k=2$ & $k=4$ & $k=8$ \\ 
        \midrule
        \multirow{ 3}{*}{
        \rotatebox[origin=c]{90}{\parbox{1cm}{Erdős-Rényi}}
        }
        & X & 0.87 & 0.89 & 0.97\\
        & Grover$^\Box$ & 0.87 & 0.90 & 0.98\\
        & Grover & 0.79 & 0.83 & 0.93\\
        \midrule
        \multirow{ 3}{*}{
        \rotatebox[origin=c]{90}{\parbox{1.4cm}{Barabási-Albert}}
        } 
        & X & 0.80 & 0.87 & 0.96\\
        & Grover$^\Box$ & 0.80 & 0.88 & 0.96\\
        & Grover & 0.74 & 0.82 & 0.92\\
        \bottomrule
        \end{tabular}
    \caption{Approximation ratios achieved for the case when $k=2^{n_k}$.}
        \label{tab:alphaP2}
\end{table}

When $k=2^{n_k}$ the equivalence relation for the problem Hamiltonian in Equation~\eqref{eq:Hij} becomes trivial, i.e. $\clr = \overbar{ \{\} } = \{(0,0), (1,1),\cdots,(k-1,k-1) \}$, and hence
\begin{equation}
\label{eq:Hii}
    \widehat H_{e} = \sum_{i = 0 }^{k-1} \ket{i}\bra{i} \otimes \ket{i}\bra{i}.
\end{equation}
By definition $\widehat H_e$ is a projection operator $P_B$ of rank $k$ onto the subspace spanned by
$B = \{ \ket{j}\otimes \ket{j}, 0\leq j < k \} $, consisting of all states with identical basis vectors in both registers.
We notice that the set $B$ is Pauli X-orbit generated as defined in~\cite{Fuchs:2025eqs}, since it can be written as
\begin{equation*}
    B = G_{n_k} \ket{0}^{\otimes 2 n_k},
\end{equation*}
i.e., the orbit of the computational basis state $\ket{0}^{\otimes 2 n_k}$ under the group action of $G_{n_k} = \langle \{X_i\otimes X_i\}_{i=0}^{n_k-1}\rangle$.
Here, $\langle S\rangle$ denotes the subgroup generated by a subset $S$ of the Pauli group.
To construct the phase separating operator for our concrete case we follow the proof of Theorem 3 in~\cite{Fuchs:2025eqs} which is recursively giving us the following permutation matrix in each step, namely
$M_j = C^j_1 X_{j+n_k}$, which leaves the all zero state unchanged and maps the state $\ket{j}\otimes\ket{j}$ to $\ket{j}\ket{0}$.
Overall this leads to
\begin{equation*}
    M P_{B} M^{\dag} = I^{n_k} \otimes \ket{0}\bra{0}^{n_k},
    \quad
    M = 
    \prod_{i=0}^{ n_k-1}C^i_1 X_{i+n_k},
\end{equation*}
resulting in the following circuit
\begin{equation}
    e^{it \widehat H_e} =
    \text{ \input{figures/Up2} }.
\label{eq:ep2}
\end{equation}
The number of controlled gates required for realizing the mixer and phase separating operators are as follows.
\begin{itemize}
    \item 
Realizing $e^{-it \widehat H_e}$ through Equation~\eqref{eq:ep2}
one needs 
$2 n_k$ $CX$ gates and 1 $C^{n_k-1}Ph$ gate.
Alternatively, one can also do a direct decomposition of $\widehat H_e$ in the Pauli basis and perform Pauli evolution gates with this.
For $k=2,4,8$, the first approach needs 2$CX$, (4$CX$ and 1 $CPh$), and  (6$CX$ and 1$C^2Ph$), whereas the Pauli evolution needs 2$CX$, 10$CX$, 34$CX$ per gate, respectively.
Thus, for these cases it is better to realize the phase separation with the circuit given in Equation~\eqref{eq:ep2} and not through Pauli evolution.
\item
The X mixer does not require any controlled gates, the Grover mixer needs 1 $C^{|V|n_k-1}P$ gate, and the Grover$^{\Box^V}$ needs $|V|$ $C^{n_k-1}P$ gates, where the box product is defined in Equation~\eqref{eq:boxproduct}.
\end{itemize}

We numerically test the performance of the resulting QAOA algorithm on the same instances as in~\cite{fuchs2021efficient}, given in Appendix~\ref{appendix:graphs}.
Table~\ref{tab:alphaP2} shows that the achieved approximation ratios are very good, for all cases.
An overview of the energy/approximation ratio landscapes are provided in the Appendix in Table~\ref{table:landscapesP2}.

\section{Binary encodings for \texorpdfstring{$k\neq 2^{n_k}$}{k!=2Lk} using the full Hilbert space}\label{sec:notp2fullH}
When $k$ is not a power of two, the equivalence relation for the problem Hamiltonian in Equation~\eqref{eq:Hij} becomes non-trivial.
One way to deal with this is what we will refer to as the $\leq k$-encoding where we encode the color $k$ into all states with $\ket{i}$ for $k\leq i\leq 2^{n_k}-1$.
This means we have
\begin{equation*}
    \clr^k_{< k} = \overbar{
    \{(k-1,k), (k,k+1),\cdots,(2^{n_k}-2,2^{n_k}-1)\}}.
\end{equation*}
Of course, in general there are many ways to define $k$ colors through an equivalence relation $\clr$, which influences the cardinality of the equivalence classes.
There is, however, always a way to define it such that the maximum cardinality of all equivalence classes is at most two.
In the following we will denote such a relation as $\clr^k_\text{bal}$.
This also influences the rank of the Hamiltonian $\widehat H_e$.
In the following, we will make a concrete construction of the Hamiltonians for $k\in\{3,5,6,7\}$.

\subsection{The case \texorpdfstring{$k=3$}{k=3}}
For $k=3$ we have that $\clr^3_\text{bal}=\clr^3_{< 3} = \overbar{\{(2,3)\}} $, which means that the three colors are given by $[0]=\{0\}, [1]=\{1\}, [2]=\{2,3\}$. 
The subspace of the projector is given by $B = 
\{\ket{0000}, \ket{0101}, \ket{1010}, \ket{1111}, \ket{1011}, \ket{1110}\}
$.
We can realize the phase separating operator using the low-pass controlled phase shift gate $C_{\leq 6} Ph$ in the permuted basis, as described in~\cite{Fuchs:2025eqs}.
However, to reduce the circuit complexity, we can instead divide the subspace in two Pauli X-orbit generated subspaces,
namely $B_1=\langle X_1X_3 \rangle\ket{0000}$ and $B_2=\langle X_1, X_3\rangle\ket{1010}$,
which results in the circuit
\begin{equation*}
    e^{it H_{\clr^3_{< 3}}} = 
    \input{figures/H3}.
\end{equation*}

\subsection{The case \texorpdfstring{$k=5$}{k=5}}
For $k=5$ we have that $\clr^5_{< 5} = \overbar{\{(4,5,6,7)\}} $, which means that the five colors are given by $[0]=\{0\}, [1]=\{1\}, [2]=\{2\}, [3]=\{3\}, [4]=\{4,5,6,7\}$. 
We can divide $B$ into two Pauli X-orbit generated subspaces. The first subspace containing four states is related to the first four color, namely $\ket{000},\ket{001},\ket{010}\, \ket{011}$ forming the subset $B_1=\langle X_1X_4,X_2X_5 \rangle\ket{000000}$. The second subspace contains all states related to the last degenerated color $\ket{100},\ket{101},\ket{110},\ket{111}$ forming $B_2=\langle X_1,X_2,X_4,X_5 \rangle\ket{100100}$. 
The resulting circuit is given by
\begin{equation}
\label{eq:lessthan5}
    e^{it H_{\clr^5_{< 5}}} =
        \input{figures/H5lessthank},
\end{equation}
which realizes the phase separating unitary with one single and one triple controlled phase shift gate.

Looking at the case for $k=3$ and $k=5$ it becomes clear that the structure of the circuit can be generalized for $\clr_{<k}$ with $k=2^l+1$.
The feasible subspace can always be divide into two Pauli X-orbit generated subspaces, namely $B_1=\langle X_1X_{n_k+1},\cdots,X_{l}X_{n_k+l} \rangle\ket{0\cdots0}$, related to the first $k-1$ color states characterized by 
bitstrings $x = yy$, where the first digit of $y$ is always $0$.
The second subspace related to the degenerate color is given by $B_2=\langle X_1,\cdots,X_{n_k-1},X_{n_k+1},\cdots,\allowbreak X_{2{n_k-1}} \rangle\ket{10\dots0}\ket{10\cdots0}$.
From this, we can conclude that the resulting circuit is a generalization of Equation~\eqref{eq:lessthan5}, where $e^{itH_1}$ consists of the $M_{GHZ}$ circuit where the CX between the first and $n_k$-th qubit is omitted and a phase shift gate with target on the last qubit and controls on the first and last $n_k$ (without the last) qubits.
The circuit for $e^{itH_2}$ is a phase shift gate with control on the first and target on the $n_k+1$-th qubit.

A second possibility is to encode the problem so that the cardinality of the equivalence classes is minimized. This can for instance be achieved by
$\clr^5_\text{bal} = \overbar{\{(0,1),(4,5),(6,7))\}} $. In this case the five colors are given by $[0]=\{0,1\}, [2]=\{2\}, [3]=\{3\}, [4]=\{4,5\}, [6]=\{6,7\}$ and the according Pauli X-orbit generated subspaces are
\begin{align}
\label{eq:B5bal}
    B_1&=\langle X_0 X_3,X_1 X_4,X_2 X_5 \rangle\ket{000000},\\
    B_2&= \langle X_0 X_3,X_2 X_5 \rangle\ket{000001},\\
    B_3&= \langle X_2X_5 \rangle\ket{111110}.
\end{align}
The resulting circuit for the phase separating operator is
\begin{equation}
    e^{it H_{\clr^5_\text{bal}}} =
    \text{ \tikzexternaldisable 
\begin{tikzpicture}[baseline,remember picture]
\node[scale=1] {
\begin{quantikz}[row sep={0.5cm,between origins},column sep=1ex]
     & \ctrl{3}
    \gategroup[6,steps=9,style={dashed,rounded corners,fill=blue!20, inner xsep=0pt},background, label style={label position=below, yshift=-0.5cm}]{
    $e^{i t H_1}$
    }
                   &       &       &       &               &       &       &       & \ctrl{3} \tikzmark{Da} & \ctrl{3} \tikzmark{Ca} 
    \gategroup[6,steps=7,style={dashed,rounded corners,fill=green!20, inner xsep=0pt},background, label style={label position=below, yshift=-0.5cm}]{
    $e^{i t H_2}$
    }
                                                                                                                                                        &         &       &              &       &         & \ctrl{3} &  
    \gategroup[6,steps=3,style={dashed,rounded corners,fill=red!20, inner xsep=0pt},background, label style={label position=below, yshift=-0.5cm}]{
    $e^{i t H_3}$
    }
                                                                                                                                                                                                                    & \ctrl{1}        &       &   \\
     &       & \ctrl{3} &       &       &               &       &       & \ctrl{3} &                     &                   &         & \gate{X} & \ctrl{1}        & \gate{X} &         &       &      & \ctrl{2}        &  \      &   \\
     &       &       & \ctrl{3} &       &               &       & \ctrl{3} &       &                     &                   & \targ{}    & \gate{X} & \ctrl{1}        & \gate{X} & \targ{}    &       & \ctrl{3}&              & \ctrl{3} &   \\
     & \targ{}  &       &       & \gate{X} & \ctrl{1}         & \gate{X} &       &       & \targ{} \tikzmark{Aa}  & \targ{} \tikzmark{Ba}&         & \gate{X} & \ctrl{1}        & \gate{X} &         & \targ{}  &      & \ctrl{1}        &       &   \\
     &       & \targ{}  &       & \gate{X} & \ctrl{1}         & \gate{X} &       & \targ{}  &                     &                   &         & \gate{X} & \gate{Ph(t)}& \gate{X} &         &       &      & \ctrl{1}        &       &   \\
     &       &       & \targ{}  & \gate{X} & \gate{Ph(t)} & \gate{X} & \targ{}  &       &                     &                   & \octrl{-3} &       &              &       & \octrl{-3} &       & \targ{} & \gate{Ph(t)}& \targ{}  &   
\end{quantikz}
};
\filldraw[fill=gray!20, opacity=.5, rounded corners, very thick, dashed]
    ([shift={(-1.15cm,+1.8cm)}] pic cs:Aa) -- 
    ([shift={(-0.15cm,+1.8cm)}] pic cs:Ba) -- 
    ([shift={(-0.15cm,-.2cm)}] pic cs:Ca) -- 
    ([shift={(-1.15cm,-.2cm)}] pic cs:Da) -- cycle;
\end{tikzpicture}
\tikzexternalenable, }
    \label{eq:H5balanced}
\end{equation}
where the gates in the gray box cancel to the identity.

\subsection{The case \texorpdfstring{$k=6$}{k=6}}
For $k=6$ we have that $\clr^6_{< 6} = \overbar{\{(5,6,7)\}} $, which means that the six colors are given by $[0]=\{0\}, [1]=\{1\}, [2]=\{2\}, [3]=\{3\}, [4]=\{4\}, [5]=\{5,6,7\}$. 
We can divide $B$ into the following two Pauli X-orbit generated subspaces $B_1=\langle X_0X_3,X_1X_4,X_2X_5 \rangle\ket{000000},B_2=\langle X_2X_5 \rangle\ket{110111}$,
together with $B_3=\{ \ket{110101},\ket{110111},\ket{111101},\ket{111110}\}$, which can not be generated by a Pauli X-orbit.
The resulting circuit is given by
\begin{equation*}
    e^{it H_{\clr^6_{< 6}}} = 
    \input{figures/H6lessthank}.
\end{equation*}
The circuit for the real time evolution of $H_2$ contains a Toffoli gate for realizing the basis change, since the subspace is not Pauli X-orbit generated.
In particular, the circuit for the second term $H_2$ can be derived considering the fact that if we exchange the state $\ket{111110}$ and $\ket{111111}$, using a transposition gate (in this case a 5 controlled not gate $C^5 X$), the new set $\Tilde{B_2}=\langle X_2,X_4 \rangle\ket{110101}$ becomes Pauli X-orbit generated. Since all the states in the subset $B_2\cup\Tilde{B_2}$ have the bit value one in position 0,1,3,5 we can replace the use of $C^5X$ by a Toffoli gate $CCX_{2,4\rightarrow 5}$ instead.

In order to encode the problem into a balanced way one can use $\clr^6_\text{bal} = \overbar{\{(0,1), (4,5)\}}$.
Since $\clr^6_\text{bal}\cup \overbar{\{(6,7)\}} = \clr^5_\text{bal}$
we can divide $B$ into the subspace $B_1$ and $B_2$ shown in Equation~\eqref{eq:B5bal},
where the resulting circuit is shown in Equation~\eqref{eq:H5balanced}, but the circuit for $H_3$ is omitted.

\subsection{The case \texorpdfstring{$k=7$}{k=7}}
For $k=7$ we have that $\clr^7_\text{bal}=\clr^7_{< 7} = \overbar{\{(6,7)\}} $, which means that the seven colors are given by $[0]=\{0\}, \cdots, [5]=\{5\}, [6]=\{6,7\}$. 
Since $\clr^7_\text{bal}\cup \overbar{\{(0,1),(4,5)\}} = \clr^5_\text{bal}$
we can divide $B$ into the subspace $B_1$ and $B_3$ shown in Equation~\eqref{eq:B5bal},
where the resulting circuit is shown in Equation~\eqref{eq:H5balanced}, but the circuit for $H_2$ is omitted.
Notice, that also in this case the two adjacent $C^2X_5$ gates between the circuits for $H_1$ and $H_3$ cancel.

\subsection{Resource analysis}
\begin{figure}
    \centering
    \begin{subfigure}{.64\linewidth}
        \centering
        \begin{tabular}{lll}
        \toprule
        k & \cite[Section 3.2.2]{fuchs2021efficient} & this work \\
        \midrule
        3 & $1CPh, 2( 4C^2X, 1C^2Ph), 4CX$   & $1CPh, 1C^2Ph, 2CX$          \\
        5 & $1C^2Ph, 12(4C^3X, 1C^2PhP), 6CX$ & $1CPh, 1C^3Ph, 4CX$          \\
        6 & $1C^2Ph, 6(4C^3X, 1C^2PhP), 6CX$  & $1C^2Ph, 1C^3Ph, 8CX$        \\
        7 & $1C^2Ph, 2(4C^3X, 1C^2PhP), 6CX$  & $1C^2Ph, 1C^3Ph, 2C^2X, 6CX$ \\
        \bottomrule
        \end{tabular}
        \caption{Number of controlled gates to realize the phase shift gate for one edge of the graph.}
        \label{tab:phaseresources}
    \end{subfigure}
    \begin{subfigure}{.34\linewidth}
        \centering
        \begin{tikzpicture}
    \node[scale=1.] {
        \begin{quantikz}[row sep={0.2cm,between origins},column sep=1pt]
            \lstick{$q_{i,0}$}       & \gate[5,nwires={2,3,4,5}]{N_1}\gategroup[5,steps=9,style={dashed,rounded corners,fill=green!20, inner xsep=-2pt, inner ysep=-3pt},background]{}& \qw    & \ctrl{10} & \qw      & \qw                            & \qw       & \ctrl{10}   & \qw & \gate[5,nwires={2,3,4,5}]{N_1} &\qw\\
            \lstick{} & &  &           &          &                                &           &            &  & & \\
            \lstick{\vdots\hphantom{a}}        &\qw& \qw    & \ctrl{8}  & \qw      & \qw                            & \qw       & \ctrl{8}    & \qw & \qw& \qw\\
            \lstick{}& &  &           &          &                                &           &             & &\\
            \lstick{$q_{i,L-1}$}    &\qw & \qw       & \ctrl{6}  & \qw      & \qw                            & \qw       & \ctrl{6}    & \qw & \qw & \qw\\[.5cm]
            \lstick{$q_{j,0}$}        &\gate[5,nwires={2,3,4,5}]{N_2}\gategroup[5,steps=9,style={dashed,rounded corners,fill=green!20, inner xsep=-2pt, inner ysep=-3pt},background]{}  & \qw      & \qw       & \ctrl{6} & \qw                            & \ctrl{6}  & \qw         & \qw & \gate[5,nwires={2,3,4,5}]{N_2} & \qw\\
            \lstick{}& & &            &          &                                &           &             & &  &  \\
            \lstick{\vdots\hphantom{a}}        &\qw   & \qw     & \qw       & \ctrl{4} & \qw                            & \ctrl{4}  & \qw         & \qw & \qw& \qw\\
            \lstick{}& &  &           &          &                                &           &             &  & & \\
            \lstick{$q_{j,L-1}$}    &\qw   & \qw     & \qw       & \ctrl{2} & \gate{Ph(-\theta)}& \ctrl{2}  & \qw         & \qw & \qw & \qw\\[.5cm]
            \lstick{$a_0$}          & \qw& \qw    & \targ{}\gategroup[2,steps=5,style={dashed,rounded corners,fill=red!20, inner ysep=-2pt},background]{}   & \qw      & \ctrl{-1}                      & \qw       & \targ{}     & \qw& \qw& \qw\\
            \lstick{$a_1$}    & \qw& \qw    & \qw       & \targ{}  & \ctrl{-1}                      & \targ{}   & \qw & \qw& \qw& \qw
        \end{quantikz}
    };
\end{tikzpicture}
        \vspace{-1.5\baselineskip}
        \caption{
        Phase shift gate required in previous work~\cite{fuchs2021efficient}.
        }
        \label{fig:oldsubcirc}
    \end{subfigure}
    \caption{
    The resource comparison shows that the circuits for realizing the phase separating operator $U_P$ as proposed in this work require fewer resources in terms of entangling gates as previous work~\cite{fuchs2021efficient}.
    Another advantage of the proposed method is that it does not require ancilla qubits.
    Whereas we realize the diagonal operator directly, previous work~\cite{fuchs2021efficient} utilized the circuit shown in (b) between each edge $(i,j)$. 
    For each element $\{y_1,y_2\} \in \clr^k\setminus \cup_{i,j} \{i,i\}$, one such circuit is necessary with $N_{y_i} = X^{\operatorname{bin}(y_i)}$.
    }
\label{fig:enp2}
\end{figure}
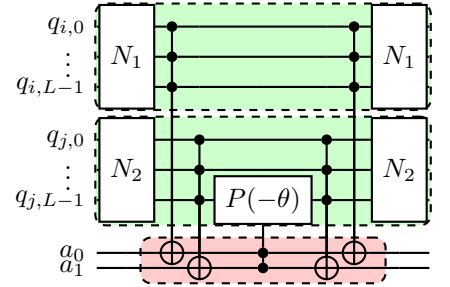

When $k$ is not a power of two, the method presented in~\cite[Section 3.2.2]{fuchs2021efficient} is not optimal in terms of resource efficiency;
for each edge of the graph, it requires the circuit shown in Equation~\eqref{eq:ep2} and
$a(a-1)$ times the circuit shown in Figure~\ref{fig:oldsubcirc}, where $a=2^{n_k}-(k-1)$.
Each of these individual circuits uses 4 $C^{n_k} X$-gates, one $C^2P$ gate and requires the use of two ancillary qubits.
Concretely, for each edge one needs $2,12,6,2,56,\cdots$ of these circuits for $k=3,5,6,7,9,\cdots$, respectively.
The method derived from Theorem \cite{Fuchs:2025eqs} on the other hand, requires much fewer entangling gates and does not require ancilla qubits. See Table~\ref{tab:phaseresources} for a comparison.

\section{Binary encodings for \texorpdfstring{$k\neq 2^{n_k}$}{k!=2Lk} using subspaces}\label{sec:notp2subspace}
When $k$ is not a power of two, there is an alternative to encoding the problem in the full Hilbert space as presented in Section~\ref{sec:notp2fullH}.
Instead, one can restrict the evolution of the QAOA to a suitably chosen subspace spanned by a set $B$ consisting of $k$ computational basis states.
Consequently, this means we need a way to prepare an initial state in a $k$-dimensional subspace for each vertex of the graph, and a mixer constrained to this subspace.
We observe that
\begin{equation*}
    \restr{H_P^\text{bin}(2^{n_k})}{\Span{B}} = H_P^\text{bin}(k),
\end{equation*}
which means we can use the circuit for the power of two case.
Overall, this approach shifts complexity from the phase separating Hamiltonian to the mixer.

\subsection{Initial state}
\begin{figure}
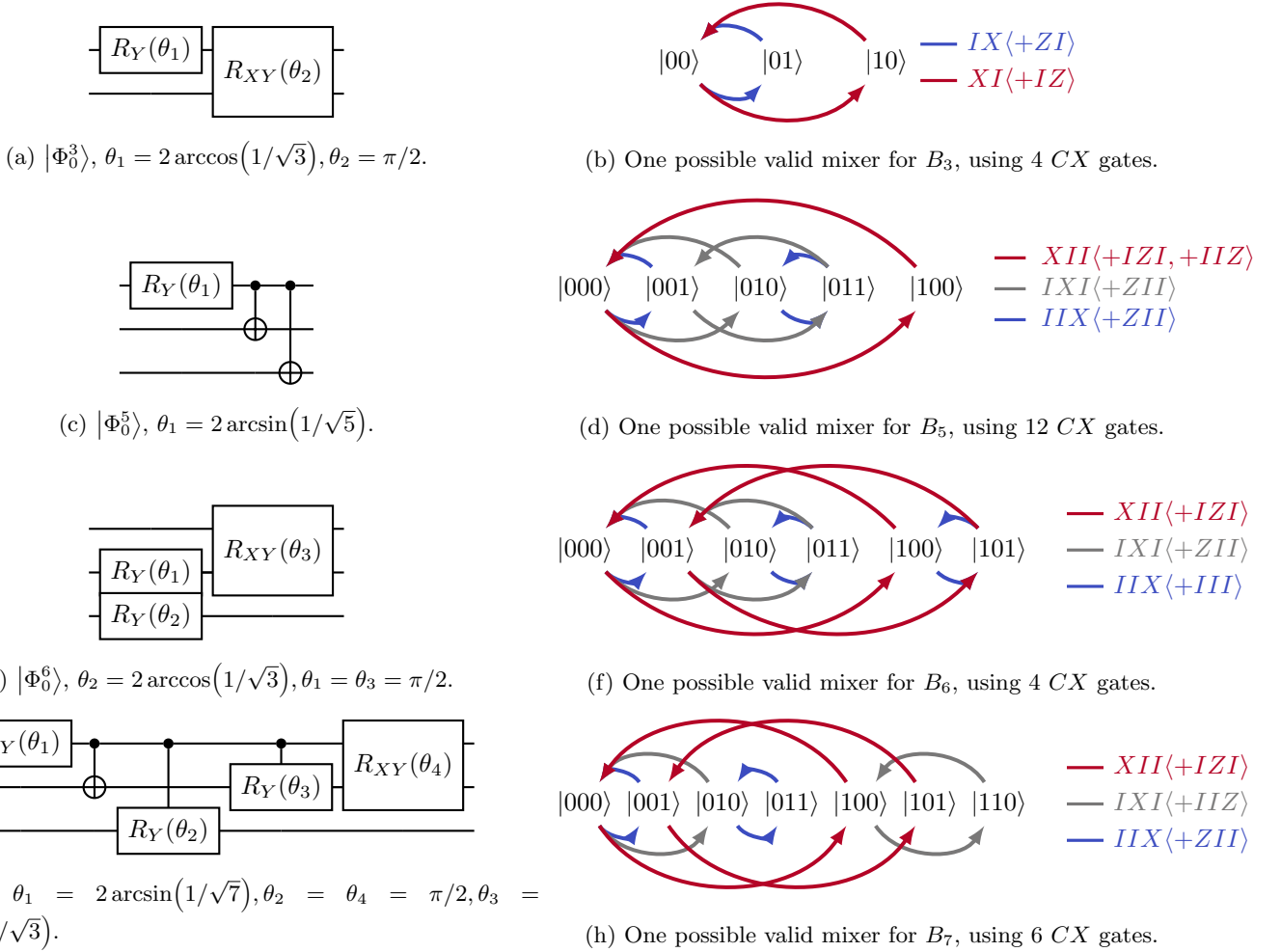

    \centering
    \begin{subfigure}[b]{.49\textwidth}
        \centering
        \input{figures/initialstateB3}
        \caption{
            $\ket{\Phi^3_0}$, $t_1 = 2 \arccos(1/\sqrt{3}), t_2=\pi/2$.
        }
    \end{subfigure}
    \begin{subfigure}[b]{.49\textwidth}
        \centering
        \begin{tikzpicture}[scale=.5]
\node[color=black,opacity=1] (00) at (0.175,0) {$\ket{00}$};
\node[color=black,opacity=1] (01) at (3.000,0) {$\ket{01}$};
\node[color=black,opacity=1] (10) at (5.825,0) {$\ket{10}$};
\Edge[,color={58.6004535,76.17308133,192.189204015},bend=-46.31,Direct,RGB](00)(01)
\Edge[,color={58.6004535,76.17308133,192.189204015},bend=-46.31,Direct,RGB](01)(00)
\Edge[,color={179.94665529,3.9668208,38.30936706},bend=-46.31,Direct,RGB](00)(10)
\Edge[,color={179.94665529,3.9668208,38.30936706},bend=-46.31,Direct,RGB](10)(00)

\def\distx{6.75}

\def\disty{.5}
\definecolor{IX}{RGB}{58.6004535,76.17308133,192.189204015};
\draw[IX,very thick] (\distx,\disty) node[right,xshift=.5cm] {$IX\langle +ZI \rangle$} -- (\distx+1,\disty);

\def\disty{-.5}
\definecolor{IX}{RGB}{179.94665529,3.9668208,38.30936706};
\draw[IX,very thick] (\distx,\disty) node[right,xshift=.5cm] {$XI\langle +IZ \rangle$} -- (\distx+1,\disty);

\end{tikzpicture}
        \caption{
        One possible valid mixer for $B_3$, using 4 $CX$ gates.
        }
    \end{subfigure}
    \begin{subfigure}[b]{.49\textwidth}
    \centering
        \input{figures/initialstateB5}
        \caption{
            $\ket{\Phi^5_0}$, $t_1 = 2 \arcsin(1/\sqrt{5})$.
        }
    \end{subfigure}
    \begin{subfigure}[b]{.49\textwidth}
        \centering
        \begin{tikzpicture}[scale=.85]
\node[color=black,opacity=1] (000) at (0.175,0) {$\ket{000}$};
\node[color=black,opacity=1] (001) at (1.587,0) {$\ket{001}$};
\node[color=black,opacity=1] (010) at (3.000,0) {$\ket{010}$};
\node[color=black,opacity=1] (011) at (4.413,0) {$\ket{011}$};
\node[color=black,opacity=1] (100) at (5.825,0) {$\ket{100}$};
\Edge[,color={58.6004535,76.17308133,192.189204015},bend=-46.31,Direct,RGB](000)(001)
\Edge[,color={58.6004535,76.17308133,192.189204015},bend=-46.31,Direct,RGB](001)(000)
\Edge[,color={58.6004535,76.17308133,192.189204015},bend=-46.31,Direct,RGB](010)(011)
\Edge[,color={58.6004535,76.17308133,192.189204015},bend=-46.31,Direct,RGB](011)(010)
\Edge[,color={121.194046947,120.416032942,119.96362781500002},bend=-46.31,Direct,RGB](000)(010)
\Edge[,color={121.194046947,120.416032942,119.96362781500002},bend=-46.31,Direct,RGB](001)(011)
\Edge[,color={121.194046947,120.416032942,119.96362781500002},bend=-46.31,Direct,RGB](010)(000)
\Edge[,color={121.194046947,120.416032942,119.96362781500002},bend=-46.31,Direct,RGB](011)(001)
\Edge[,color={179.94665529,3.9668208,38.30936706},bend=-46.31,Direct,RGB](000)(100)
\Edge[,color={179.94665529,3.9668208,38.30936706},bend=-46.31,Direct,RGB](100)(000)

\def\distx{6.75}

\def\disty{.5}
\definecolor{XII}{RGB}{179.94665529,3.9668208,38.30936706};
\draw[XII,very thick] (\distx,\disty) node[right,xshift=.5cm] {$XII\langle +IZI, +IIZ \rangle$} -- (\distx+.5,\disty);

\def\disty{0}
\definecolor{IXI}{RGB}{121.194046947,120.416032942,119.96362781500002};
\draw[IXI,very thick] (\distx,\disty) node[right,xshift=.5cm] {$IXI\langle +ZII \rangle$} -- (\distx+.5,\disty);

\def\disty{-.5}
\definecolor{IIX}{RGB}{58.6004535,76.17308133,192.189204015};
\draw[IIX,very thick] (\distx,\disty) node[right,xshift=.5cm] {$IIX\langle +ZII \rangle$} -- (\distx+.5,\disty);

\end{tikzpicture}
        \vspace{-1\baselineskip}
        \caption{
        One possible valid mixer for $B_5$, using 12 $CX$ gates.
        }
    \end{subfigure}
    \begin{subfigure}[b]{.49\textwidth}
    \centering
        \input{figures/initialstateB6}
        \caption{
            $\ket{\Phi^6_0}$, $t_2 = 2 \arccos(1/\sqrt{3}), t_1 = t_3 = \pi/2$.
        }
    \end{subfigure}
    \begin{subfigure}[b]{.49\textwidth}
        \centering
        \begin{tikzpicture}
\node[color=black,opacity=1] (000) at (0.175,0) {$\ket{000}$};
\node[color=black,opacity=1] (001) at (1.305,0) {$\ket{001}$};
\node[color=black,opacity=1] (010) at (2.435,0) {$\ket{010}$};
\node[color=black,opacity=1] (011) at (3.565,0) {$\ket{011}$};
\node[color=black,opacity=1] (100) at (4.695,0) {$\ket{100}$};
\node[color=black,opacity=1] (101) at (5.825,0) {$\ket{101}$};
\Edge[,color={58.6004535,76.17308133,192.189204015},bend=-46.31,Direct,RGB](000)(001)
\Edge[,color={58.6004535,76.17308133,192.189204015},bend=-46.31,Direct,RGB](001)(000)
\Edge[,color={58.6004535,76.17308133,192.189204015},bend=-46.31,Direct,RGB](010)(011)
\Edge[,color={58.6004535,76.17308133,192.189204015},bend=-46.31,Direct,RGB](011)(010)
\Edge[,color={58.6004535,76.17308133,192.189204015},bend=-46.31,Direct,RGB](100)(101)
\Edge[,color={58.6004535,76.17308133,192.189204015},bend=-46.31,Direct,RGB](101)(100)
\Edge[,color={121.194046947,120.416032942,119.96362781500002},bend=-46.31,Direct,RGB](000)(010)
\Edge[,color={121.194046947,120.416032942,119.96362781500002},bend=-46.31,Direct,RGB](001)(011)
\Edge[,color={121.194046947,120.416032942,119.96362781500002},bend=-46.31,Direct,RGB](010)(000)
\Edge[,color={121.194046947,120.416032942,119.96362781500002},bend=-46.31,Direct,RGB](011)(001)
\Edge[,color={179.94665529,3.9668208,38.30936706},bend=-46.31,Direct,RGB](000)(100)
\Edge[,color={179.94665529,3.9668208,38.30936706},bend=-46.31,Direct,RGB](001)(101)
\Edge[,color={179.94665529,3.9668208,38.30936706},bend=-46.31,Direct,RGB](100)(000)
\Edge[,color={179.94665529,3.9668208,38.30936706},bend=-46.31,Direct,RGB](101)(001)

\def\distx{6.75}

\def\disty{.5}
\definecolor{XII}{RGB}{179.94665529,3.9668208,38.30936706};
\draw[XII,very thick] (\distx,\disty) node[right,xshift=.5cm] {$XII\langle +IZI \rangle$} -- (\distx+.5,\disty);

\def\disty{0}
\definecolor{IXI}{RGB}{121.194046947,120.416032942,119.96362781500002};
\draw[IXI,very thick] (\distx,\disty) node[right,xshift=.5cm] {$IXI\langle +ZII \rangle$} -- (\distx+.5,\disty);

\def\disty{-.5}
\definecolor{IIX}{RGB}{58.6004535,76.17308133,192.189204015};
\draw[IIX,very thick] (\distx,\disty) node[right,xshift=.5cm] {$IIX\langle +III \rangle$} -- (\distx+.5,\disty);
\end{tikzpicture}
        \vspace{-1\baselineskip}
        \caption{
        One possible valid mixer for $B_6$, using 4 $CX$ gates.
        }
    \end{subfigure}
    \begin{subfigure}[b]{.49\textwidth}
    \centering
        \input{figures/initialstateB7}
        \caption{
            $\ket{\Phi^7_0}$, $t_1 = 2 \arcsin(1/\sqrt{7}), t_2 = t_4 = \pi/2, t_3 = 2 \arccos(1/\sqrt{3})$.
        }
    \end{subfigure}
    \begin{subfigure}[b]{.49\textwidth}
        \centering
        \begin{tikzpicture}
\node[color=black,opacity=1] (000) at (0.175,0) {$\ket{000}$};
\node[color=black,opacity=1] (001) at (1.117,0) {$\ket{001}$};
\node[color=black,opacity=1] (010) at (2.058,0) {$\ket{010}$};
\node[color=black,opacity=1] (011) at (3.000,0) {$\ket{011}$};
\node[color=black,opacity=1] (100) at (3.942,0) {$\ket{100}$};
\node[color=black,opacity=1] (101) at (4.883,0) {$\ket{101}$};
\node[color=black,opacity=1] (110) at (5.825,0) {$\ket{110}$};
\Edge[,color={58.6004535,76.17308133,192.189204015},bend=-56.31,Direct,RGB](000)(001)
\Edge[,color={58.6004535,76.17308133,192.189204015},bend=-56.31,Direct,RGB](001)(000)
\Edge[,color={58.6004535,76.17308133,192.189204015},bend=-56.31,Direct,RGB](010)(011)
\Edge[,color={58.6004535,76.17308133,192.189204015},bend=-56.31,Direct,RGB](011)(010)
\Edge[,color={121.194046947,120.416032942,119.96362781500002},bend=-56.31,Direct,RGB](000)(010)
\Edge[,color={121.194046947,120.416032942,119.96362781500002},bend=-56.31,Direct,RGB](010)(000)
\Edge[,color={121.194046947,120.416032942,119.96362781500002},bend=-56.31,Direct,RGB](100)(110)
\Edge[,color={121.194046947,120.416032942,119.96362781500002},bend=-56.31,Direct,RGB](110)(100)
\Edge[,color={179.94665529,3.9668208,38.30936706},bend=-56.31,Direct,RGB](000)(100)
\Edge[,color={179.94665529,3.9668208,38.30936706},bend=-56.31,Direct,RGB](001)(101)
\Edge[,color={179.94665529,3.9668208,38.30936706},bend=-56.31,Direct,RGB](100)(000)
\Edge[,color={179.94665529,3.9668208,38.30936706},bend=-56.31,Direct,RGB](101)(001)

\def\distx{6.75}

\def\disty{.5}
\definecolor{XII}{RGB}{179.94665529,3.9668208,38.30936706};
\draw[XII,very thick] (\distx,\disty) node[right,xshift=.5cm] {$XII\langle +IZI \rangle$} -- (\distx+.5,\disty);

\def\disty{0}
\definecolor{IXI}{RGB}{121.194046947,120.416032942,119.96362781500002};
\draw[IXI,very thick] (\distx,\disty) node[right,xshift=.5cm] {$IXI\langle +IIZ \rangle$} -- (\distx+.5,\disty);

\def\disty{-.5}
\definecolor{IIX}{RGB}{58.6004535,76.17308133,192.189204015};
\draw[IIX,very thick] (\distx,\disty) node[right,xshift=.5cm] {$IIX\langle +ZII \rangle$} -- (\distx+.5,\disty);
\end{tikzpicture}
        \vspace{-1\baselineskip}
        \caption{
        One possible valid mixer for $B_7$, using 6 $CX$ gates.
        }
    \end{subfigure}
    \caption{Initial state preparation and mixer for $k=3,5,6,7$ when restricting to a suitably chosen subspace.
        We use the notation 
        $R_{P}(t) = \exp \left(-i\frac {t}{2}p\right)$, $P\in\{X,Y,Z\}$, and 
        $R_{XY}(t) = \exp \left(-i\frac {t}{4}(X\otimes X+Y\otimes Y)\right)$.
    }
    \label{fig:subspaceCircuits}
\end{figure}

For the \maxkcut{k} problem, one possibility is to define the subspace for each vertex through $B_k = \{\ket{i} \ | \ 0 \leq i \leq k-1\}$.
Since the feasible bit-strings admit a Cartesian product structure, the feasible subspace becomes a tensor product of the form
\begin{equation*}
    B_k^1\otimes \cdots \otimes B_k^{|V|}.
\end{equation*}
An initial state can be prepared as the uniform superposition of all feasible basis states, i.e. 
\begin{equation*}
    \ket{\Phi^{<k}_0}^{\otimes^{|V|}}, \text{ where } \ket{\Phi^{<k}_0}=\frac{1}{\sqrt{k}}\sum_{i=0}^{k-1}\ket{i}.
\end{equation*}
This can in general be efficiently prepared with a gate complexity and circuit depth of only $\Order{\log_2(k)}$~\cite{Shukla_2024}. One way to realize the initial state for $k\in{3,5,6,7}$ is shown in Figure~\ref{fig:subspaceCircuits}.

\subsection{Constrained Mixer}
Given a feasible subspace $\Span{B}\subset \mathcal{H}$ a mixer $U_M$ is then called valid~\cite{hadfield2019quantum} if it 
\textit{preserves the feasible subspace}, i.e.
\begin{equation*}
    U_M(\beta)\ket{v} \in \Span{B}, \quad \forall \ket{v} \in \Span{B}, \forall \beta\in\R,
\end{equation*}
and if it \textit{provides transitions between all pairs of feasible states}, i.e. for each pair of computational basis states $\ket{x}, \ket{y} \in B$ there exist
$\beta^*\in\R$ and $r \in \N\cup\{0\}$, such that 
\begin{equation*}
    |\bra{x} U_M^r(\beta^*) \ket{y}| > 0.
\end{equation*}
Apart from the Grover mixer, we consider here constraint preserving mixers in the stabilizer form~\cite{fuchs2023LX}, in which case the Hamiltonian can be written as
\begin{equation*}
    H_M = \sum_\alpha c_\alpha \lX^{\alpha} \prod\nolimits_{V_{\alpha}}, 
\end{equation*}
where $\alpha\in\{0,1\}^n\setminus\{0\}^n$ is a multi-index, 
$\lX^\alpha  = X_1^{\alpha_1} \cdots X_n^{\alpha_n}\in \{I,X\}^n\setminus\{I\}^n$, and $\prod\nolimits_{V_{\alpha}}$ is a projection onto $V_{\alpha}$.
Furthermore, $V_\alpha$ is a subspace fulfilling  
$\ket{x}\in V_{\alpha} \Rightarrow \lX^\alpha \ket{x} \in V_{\alpha}$.
The free parameter $c_\alpha\in\{0,1\}$ must be chosen such that we have a valid mixer.
Let $\{ S_1, \cdots, S_l \}$ be a minimal generating set of the stabilizer group $S=\langle S_1, \cdots, S_l \rangle$ defining the code subspace $C(S)$.
We can choose the projection operator to be of the form
\begin{equation*}
    \prod\nolimits_{C(S)} = \frac{1}{2^k} \sum_{S\in\langle S_1, \cdots, S_k\rangle} S.
\end{equation*}
For more details we refer the reader to~\cite{fuchs2023LX}.
In the following, we will use the notation
\begin{equation*}
\lX \langle S_1, \cdots, S_k\rangle 
    \coloneqq \frac{1}{2^k} \sum_{S \in \langle S_1, \cdots, S_k\rangle}\lX S.
\end{equation*}

Given the independence of the different color constraints for each vertex of the problem, we can then define the following LX-mixer Hamiltonian
\begin{equation*}
    H_{M}=G^{\Box^{|V|}} = G\Box G \cdots \Box G,
\end{equation*}
where the Cartesian or Box product is defined as 
\begin{equation}
\label{eq:boxproduct}
    G\Box H=G\otimes\mathbb{I}+\mathbb{I}\otimes H.
\end{equation}
It can be shown that if $G$ is a valid mixer, so is $H_M$ constructed in this way~\cite{fuchs2023LX}.
Valid mixers with optimal cost in terms of $CX$ gates are shown in Figure~\ref{fig:subspaceCircuits} for $k\in\{3,5,6,7\}$.
We note, that this choice is not unique and other mixers are possible as well.

\section{Simulations}
We perform numerical simulations for an unweighted Erdős-Rényi graph and a weighted Barabási-Albert graph with 10 vertices, shown in Appendix~\ref{appendix:graphs}.
Since the case when $k$ is a power of two is presented in Section~\ref{sec:poweroftwo}, we will focus here on evaluating the methods presented in Sections~\ref{sec:notp2fullH} and \ref{sec:notp2subspace}.
Using an ideal simulator we compare the results for the binary encoding with the standard $X$-mixer and different choices of equivalence relations to encode the colors, and the subspace encoding with different choices of mixers.
The resulting energy landscapes are provided in Appendix~\ref{appendix:graphs}.
As expected, the landscapes are periodic in $\beta$ when using the $X$- and Grover-mixer, whereas the landscape associated to the LX-mixer is not in general.
The resulting approximation ratios are shown in Table~\ref{table:approxnp2}.
Overall, the subspace-encoding using the Grover$^\Box$ mixer achieves the highest values for these two graphs, although with only a small distance to the LX-mixer.
We can also see, that the balancing the bin sizes for the different colors has a very positive effect on the approximation ratio, when using the full Hilbert space and the X-mixer.
The effect of increasing the depth is shown in Table~\ref{table:depthnp2}, which is exemplified for $k=6$.
We remark that the behaviour for $k=3,5,6,$ and $7$ were very similar.
Overall, the Grover$^\Box$ achieved the best convergence for these two graphs, irrespective of if the full Hilbert space, or the subspace method was used.

We would like to note, however, that one should make a careful consideration of the averaged achieved approximation ratio with respect to the required circuit depth.
For the encoding in the full Hilbert space, one can use the X-mixer, but the the phase-separation operator requires a deeper circuit.
For the encoding into a subspace, one can use the phase-separation operator for the power of two case (leading to a more shallow circuit), but the circuit for the constrained preserving mixer is deeper.
An indication of the cost is given in Table~\ref{table:costnp2}, where we used the compiler from qiskit, assuming full connectivity.
Note that for a sparse graph $|E| = \mathcal{O}(|V|)$ and for a fully connected graph $ |E| = \Theta(|V|^2)$.
Therefore, for graphs with sufficient connectivity, the subspace encoding uses fewer resources than the encoding into the full Hilbert space.

\begin{table}
    \centering
        \begin{subtable}[t]{1\textwidth}
            \centering
            \begin{tabular}{llllrrrr}
                \toprule
                & mixer & $\ket{\phi_0}$ &$\clr$& $k=3$ & $k=5$ & $k=6$ & $k=7$ \\ 
                \midrule
                \multirow{ 6}{*}{
                \rotatebox[origin=c]{90}{\parbox{1.5cm}{Erdős-Rényi}}
                } 
                & X & $\ket{+}$  &$\clr^k_{< k}$                        & 0.80& 0.86& 0.92& 0.95\\
                & Grover$^\Box$ & $\ket{+}$  &$\clr^k_{< k}$            & 0.81& 0.90&  0.94& 0.96 \\
                & X &  $\ket{+}$     &$\clr^k_\text{bal}$               & - & 0.90& 0.94& - \\
                & Grover$^\Box$ & $\ket{+}$  &$\clr^k_\text{bal}$       & - & 0.93& 0.95& - \\
        
                & \cellcolor{lightgray} LX            & \cellcolor{lightgray} $\ket{\Phi^{<k}_0}$ &\cellcolor{lightgray} - &\cellcolor{lightgray}  0.80&\cellcolor{lightgray}  0.91&\cellcolor{lightgray}  0.94&\cellcolor{lightgray}  0.94\\
                & \cellcolor{lightgray} Grover$^\Box$ & \cellcolor{lightgray} $\ket{\Phi^{<k}_0}$ &\cellcolor{lightgray} - &\cellcolor{lightgray}  0.83&\cellcolor{lightgray}  0.94&\cellcolor{lightgray}  0.96&\cellcolor{lightgray}  0.97\\
                & \cellcolor{lightgray} Grover        & \cellcolor{lightgray} $\ket{\Phi^{<k}_0}$ &\cellcolor{lightgray} - &\cellcolor{lightgray}  0.75&\cellcolor{lightgray}  0.87&\cellcolor{lightgray}  0.90&\cellcolor{lightgray}  0.92\\
                \midrule
                \multirow{ 6}{*}{
                \rotatebox[origin=c]{90}{\parbox{1.5cm}{Barabási-Albert}}
                } 
                & X & $\ket{+}$  &$\clr^k_{< k}$                               & 0.79& 0.84& 0.90& 0.93\\
                & Grover$^\Box$ & $\ket{+}$  &$\clr^k_{< k}$                   & 0.79& 0.88& 0.92& 0.95 \\
                & X & $\ket{+}$  &$\clr^k_\text{bal}$                         & - & 0.88& 0.92& - \\
                & Grover$^\Box$ & $\ket{+}$  &$\clr^k_\text{bal}$        & - & 0.91& 0.93& - \\
                & \cellcolor{lightgray} LX            &\cellcolor{lightgray}  $\ket{\Phi^{<k}_0}$  &\cellcolor{lightgray} - &\cellcolor{lightgray}  0.79&\cellcolor{lightgray}  0.89&\cellcolor{lightgray}  0.93&\cellcolor{lightgray}  0.93\\
                & \cellcolor{lightgray} Grover$^\Box$ &\cellcolor{lightgray}  $\ket{\Phi^{<k}_0}$  &\cellcolor{lightgray} - &\cellcolor{lightgray}  0.82&\cellcolor{lightgray}  0.92&\cellcolor{lightgray}  0.94&\cellcolor{lightgray}  0.95\\
                & \cellcolor{lightgray} Grover        &\cellcolor{lightgray}   $\ket{\Phi^{<k}_0}$ &\cellcolor{lightgray} - &\cellcolor{lightgray}  0.75&\cellcolor{lightgray}  0.86&\cellcolor{lightgray}  0.89&\cellcolor{lightgray}  0.91\\
                \bottomrule
            \end{tabular}
            \caption{Approximation ratios achieved for the graphs shown in Appendix~\ref{appendix:graphs}.
            The light gray color indicates methods using subspaces as described in Section~\ref{sec:notp2subspace},
            whereas the rest indicates methods using the full Hilbert space as described in Section~\ref{sec:notp2fullH}.
            }
            \label{table:approxnp2}
        \end{subtable}
        
        \begin{subtable}[t]{1\textwidth}
        \centering
\begin{tikzpicture}

\definecolor{gray127}{RGB}{127,127,127}
\definecolor{gray1408675}{RGB}{140,86,75}
\definecolor{green4416044}{RGB}{44,160,44}
\definecolor{red2143940}{RGB}{214,39,40}
\definecolor{red25512714}{RGB}{255,127,14}
\definecolor{silver176}{RGB}{176,176,176}
\definecolor{silver204}{RGB}{204,204,204}
\definecolor{silver227119194}{RGB}{227,119,194}
\definecolor{teal31119180}{RGB}{31,119,180}

\begin{axis}[
scale=.7,
ylabel={$ \alpha $},
xlabel={$ p $},
tick align=outside,
tick pos=left,
title={Erdős-Rényi $k=6$},
axis lines=left,
x grid style={silver176},
xmin=0.8, xmax=5.2,
xtick style={color=black},
y grid style={silver176},
ymin=0.85, ymax=1,
ytick style={color=black}
]

\addplot [semithick, black, mark=square, mark size=3, mark options={solid}, forget plot]
table {%
1 1
2 1
3 1
4 1
5 1
};

\addplot [semithick, teal31119180, mark=x, mark size=3, mark options={solid}]
table {%
1 0.917188124999996
2 0.956655625000005
3 0.979811250000009
4 0.991725000000009
5 0.993869999999996
};
\addplot [semithick, red25512714, mark=x, mark size=3, mark options={solid}]
table {%
1 0.935320625000006
2 0.979798124999996
3 0.992446874999994
4 0.996855624999991
5 0.998540625000001
};
\addplot [semithick, green4416044, mark=x, mark size=3, mark options={solid}]
table {%
1 0.932576874999998
2 0.970808125000003
3 0.989001874999997
4 0.994148750000004
5 0.996206875000004
};
\addplot [semithick, red2143940, mark=x, mark size=3, mark options={solid}]
table {%
1 0.947010000000005
2 0.984504374999999
3 0.995199999999998
4 0.997851250000005
5 0.998784999999998
};
\addplot [semithick, gray1408675, dashed, mark=*, mark size=3, mark options={solid}]
table {%
1 0.941229999999993
2 0.968921875000007
3 0.984452499999999
4 0.989440624999998
5 0.992155000000004
};
\addplot [semithick, silver227119194, dashed, mark=*, mark size=3, mark options={solid}]
table {%
1 0.896328124999998
2 0.923029999999996
3 0.943827499999998
4 0.955649375000005
5 0.966184999999998
};
\addplot [semithick, gray127, dashed, mark=*, mark size=3, mark options={solid}]
table {%
1 0.957164375000001
2 0.989899375000002
3 0.997768125
4 0.999187499999998
5 0.999481875000003
};

\end{axis}

\end{tikzpicture}
\begin{tikzpicture}

\definecolor{gray127}{RGB}{127,127,127}
\definecolor{gray1408675}{RGB}{140,86,75}
\definecolor{green4416044}{RGB}{44,160,44}
\definecolor{red2143940}{RGB}{214,39,40}
\definecolor{red25512714}{RGB}{255,127,14}
\definecolor{silver176}{RGB}{176,176,176}
\definecolor{silver204}{RGB}{204,204,204}
\definecolor{silver227119194}{RGB}{227,119,194}
\definecolor{teal31119180}{RGB}{31,119,180}

\begin{axis}[
scale=.7,
legend cell align={left},
legend style={
  fill opacity=0.8,
  draw opacity=1,
  text opacity=1,
  at={(-1.0,0.03)},
  anchor=south west,
  draw=silver204
},
ylabel={$ \alpha $},
xlabel={$ p $},
tick align=outside,
tick pos=left,
title={Barabási-Albert $k=6$},
axis lines=left,
x grid style={silver176},
xmin=0.8, xmax=5.2,
xtick style={color=black},
y grid style={silver176},
ymin=0.85, ymax=1,
ytick style={color=black}
]

\addplot [semithick, black, mark=square, mark size=3, mark options={solid}]
table {%
1 1
2 1
3 1
4 1
5 1
};
\addlegendentry{Optimal}

\addplot [semithick, gray1408675, dashed, mark=*, mark size=3, mark options={solid}]
table {%
1 0.924997626791932
2 0.945964526574121
3 0.957433031287903
4 0.963023880633211
5 0.969580956422716
};
\addlegendentry{LX}
\addplot [semithick, silver227119194, dashed, mark=*, mark size=3, mark options={solid}]
table {%
1 0.890430898634199
2 0.91165094586268
3 0.932070963562271
4 0.943633390358737
5 0.952684508992378
};
\addlegendentry{Grover}
\addplot [semithick, gray127, dashed, mark=*, mark size=3, mark options={solid}]
table {%
1 0.939003393802365
2 0.96830087206736
3 0.979031760434037
4 0.98381694056834
5 0.987331137500412
};
\addlegendentry{$\text{Grover}^\Box$}

\addplot [semithick, teal31119180, mark=x, mark size=3, mark options={solid}]
table {%
1 0.901530381488679
2 0.928151122575185
3 0.94733482134885
4 0.960836046674154
5 0.971845977848825
};
\addlegendentry{$clr^k_{< k}$, X}
\addplot [semithick, red25512714, mark=x, mark size=3, mark options={solid}]
table {%
1 0.917706880925
2 0.956442485374912
3 0.971419723056338
4 0.979903703291293
5 0.984318423944705
};
\addlegendentry{$clr^k_{< k}$, $\text{Grover}^\Box$}
\addplot [semithick, green4416044, mark=x, mark size=3, mark options={solid}]
table {%
1 0.916494903584042
2 0.940158054632576
3 0.953266202082535
4 0.963537611198902
5 0.971122026020821
};
\addlegendentry{$clr^k_\text{bal}$, X}
\addplot [semithick, red2143940, mark=x, mark size=3, mark options={solid}]
table {%
1 0.928815724283122
2 0.9620340063468
3 0.975155641067342
4 0.981048475437358
5 0.985396232654377
};
\addlegendentry{$clr^k_\text{bal}$, $\text{Grover}^\Box$}

\end{axis}

\end{tikzpicture}
            \caption{Approximation ratio for $k=6$ for the different methods as the depth $p$ is increased
            employing the interpolation-based heuristic described in~\cite{Zhou2020}.
            The subspace methods constrained to subspaces use dashed lines with circles as makers, and the methods using the full Hilbert space use solid lines with "x" as markers.
            }
        \label{table:depthnp2}
        \end{subtable}
    \begin{subtable}[t]{1\textwidth}
        \centering
        \begin{tabular}{lllrrrr}
        \toprule
        &&& $k=3$ & $k=5$ & $k=6$ & $k=7$ \\ 
        \midrule
        \rowcolor{verylightgray}
        X& $\ket{+}$& $H^{<k,\text{old}}_P(k)$     & 70$|E|$& 782$|E|$& 398$|E|$& 142$|E|$\\
        \rowcolor{verylightgray}
        X& $\ket{+}$& $H^{<k, \text{Pauli}}_P(k)$  & 18$|E|$& 66$|E|$& 258$|E|$& 130$|E|$\\
        X& $\ket{+}$& $H^{<k}_P(k)$           & 12$|E|$& 26$|E|$& 90$|E|$& 58$|E|$\\
        \rowcolor{verylightgray}
        X& $\ket{+}$& $H^{\text{bal}, \text{old}}_P(k)$  & - & 398$|E|$& 270$|E|$& -\\
        \rowcolor{verylightgray}
        X& $\ket{+}$& $H^{\text{bal}, \text{Pauli}}_P(k)$  & - & 130$|E|$& 66$|E|$& -\\
        X& $\ket{+}$& $H^\text{bal}_P(k)$     & - & 82$|E|$& 36$|E|$& -\\
        LX& $\ket{\Phi^{<k}_0}$& $H_P(2^{n_k})$            & (2+4)$|V|$ + 6$|E|$& (2+12)$|V|$ + 14$|E|$& (2+4)$|V|$ + 14$|E|$& (5+6)$|V|$ + 14$|E|$\\
        Grover$^\Box$& $\ket{\Phi^{<k}_0}$& $H_P(2^{n_k})$ & (2+6)$|V|$ + 6$|E|$& (2+12)$|V|$ + 14$|E|$& (2+12)$|V|$ + 14$|E|$& (5+18)$|V|$ + 14$|E|$\\
        \bottomrule
        \end{tabular}
        \caption{
        Cost in terms of CX gates, when using the qiskit compiler.
        Methods from previous work is marked with light gray background:
        The line with $H^\text{old}_P(k)$ is the Hamiltonian presented in~\cite{fuchs2021efficient},
        and the $H^\text{Pauli}_P(k)$ decomposes the Hamiltonian in the Pauli basis.
        Terms related to the phase separating Hamiltonian scale with the number of edges and terms related to the mixer scale with the number of vertices.
        Notice that by encoding the problem in a specific subspace, rather than the entire Hilbert space, the complexity is shifted away from the phase-separating Hamiltonian and instead transferred to the mixer.
        }
        \label{table:costnp2}
    \end{subtable}
        \caption{
        Comparison of different encoding schemes when $k$ is not a power of two.
            The X-mixer uses the encoding of the full Hilbert space presented in Section~\ref{sec:notp2fullH}, whereas the LX- and Grover mixers encode the problem in a subspace as described in Section~\ref{sec:notp2subspace}.
        }
        \label{table:np2}
\end{table}

\section{Availability of Data and Code}
All data, e.g. graphs, and the python/jupyter notebook source code of the \maxkcut{k}-implementation using QAOA for reproducing the results obtained in this article are available at \url{https://github.com/OpenQuantumComputing/QAOA/}.
The code has been thoroughly developed with unit tests.

\section{Author contributions}
Franz G. Fuchs formulated the concept, developed the methodology, wrote the software, made the formal analysis and investigation, wrote the article, made the visualizations.
Ruben Pariente Bassa developed the methodology, wrote part of the software, made the analysis, contributed to the article.
Frida Lien wrote part of the software, contributed to the article.

\section{Acknowledgment}
We would like to thank for funding of the work by the Research Council of Norway through project number 332023. The authors wish to thank Terje Nilsen at Kongsberg Discovery for the access to an H100 GPU for the simulations.

\section{Conclusion}\label{sec:conclusion}
This study presents advancements in encoding the \maxkcut{k} problem into qubit systems, with a focus on cases where $k$ is not a power of two.
We present new ways of encoding that significantly reduce the resource requirements compared to previous approaches.
Balanced color sets and constrained mixers demonstrate notable improvements in optimization landscapes and approximation ratios.
In the future we plan to do a careful analysis of which ansatz is preferable when $k$ is not a power of two, using a suit of graph instances.
The analysis should take into account the overall circuit depth, the achieved approximation ratio for increasing depth size, as well as if the landscape has barren plateaus or many ``bad'' local minima.

\bibliography{references}

\begin{appendix}
\section{Graph instances}\label{appendix:graphs}
\begin{figure}[ht]
    \centering
    \begin{subfigure}[b]{.49\textwidth}
    \centering
        \includegraphics[width=.65\linewidth]{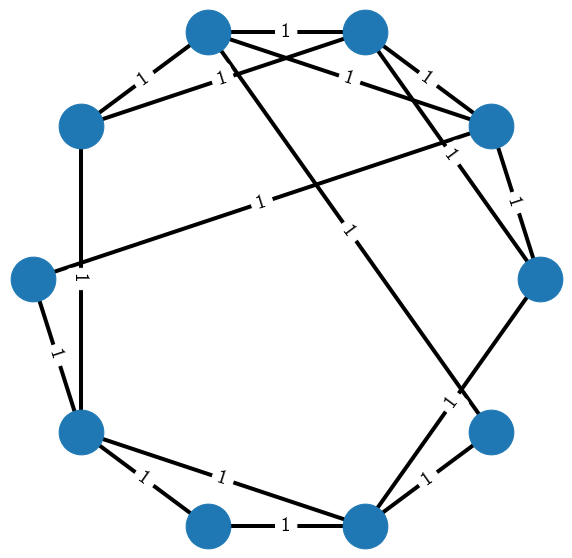}
        \caption{Instance of the unweighted Erdős-Rényi graph.}
    \end{subfigure}
    \begin{subfigure}[b]{.49\textwidth}
    \centering
        \includegraphics[width=.65\linewidth]{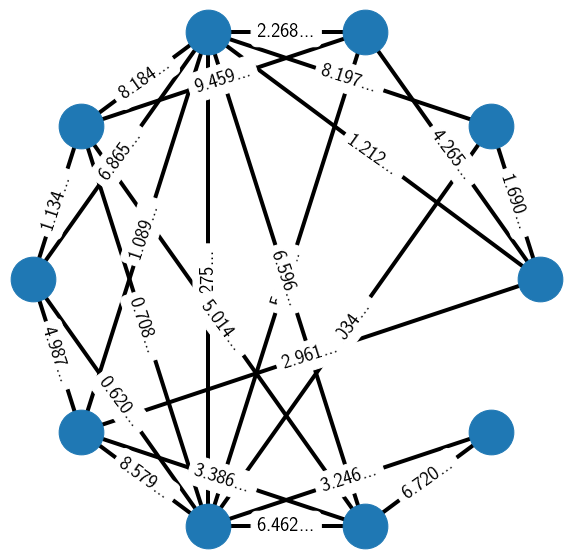}
        \caption{Instance of the weighted Barabási-Albert graph.}
    \end{subfigure}
    \caption{Graph instances used in this paper.}
    \label{fig:graph_instances}
\end{figure}

\newpage
\section{Landscapes for the approximation ratio at \texorpdfstring{$p=1$}{p1}}

\begingroup
\setlength{\tabcolsep}{0pt}
\begin{table}[ht]\sffamily
    \centering
    \begin{tabular}{llccc}
    \toprule
     & mixer & $k=2$ & $k=4$ & $k=8$ \\ 
    \midrule
    \multirow{2}{*}[-.75cm]{
    \rotatebox[origin=c]{90}{\parbox{1cm}{Barabási-Albert}}
    }&  X &
    \imshow{\datfilename{BarabasiAlbert}{2}{True}{False}{False}{all}{X}}{0.3}
    &
    \imshow{\datfilename{BarabasiAlbert}{4}{True}{False}{False}{all}{X}}{0.3}
     & 
    \imshow{\datfilename{BarabasiAlbert}{8}{True}{False}{False}{all}{X}}{0.3}
    \\
     &  Grover$^\Box$ &
    \imshow{\datfilename{BarabasiAlbert}{2}{True}{False}{False}{all}{Grovertensorized}}{0.3}
    &
    \imshow{\datfilename{BarabasiAlbert}{4}{True}{False}{False}{all}{Grovertensorized}}{0.3}
     & 
    \imshow{\datfilename{BarabasiAlbert}{8}{True}{False}{False}{all}{Grovertensorized}}{0.3}
    \\
     &  Grover &
    \imshow{\datfilename{BarabasiAlbert}{2}{True}{False}{False}{all}{Grover}}{0.3}
    &
    \imshow{\datfilename{BarabasiAlbert}{4}{True}{False}{False}{all}{Grover}}{0.3}
     & 
    \imshow{\datfilename{BarabasiAlbert}{8}{True}{False}{False}{all}{Grover}}{0.3}
    \\
    \midrule
    \multirow{2}{*}[-.75cm]{
    \rotatebox[origin=c]{90}{\parbox{1cm}{Erdős-Rényi}}
    }&  X &
    \imshow{\datfilename{ErdosRenyi}{2}{True}{False}{False}{all}{X}}{0.3}
    &
    \imshow{\datfilename{ErdosRenyi}{4}{True}{False}{False}{all}{X}}{0.3}
     & 
    \imshow{\datfilename{ErdosRenyi}{8}{True}{False}{False}{all}{X}}{0.3}
    \\
     &  Grover$^\Box$ &
    \imshow{\datfilename{ErdosRenyi}{2}{True}{False}{False}{all}{Grovertensorized}}{0.3}
    &
    \imshow{\datfilename{ErdosRenyi}{4}{True}{False}{False}{all}{Grovertensorized}}{0.3}
     & 
    \imshow{\datfilename{ErdosRenyi}{8}{True}{False}{False}{all}{Grovertensorized}}{0.3}
    \\
     &  Grover &
    \imshow{\datfilename{ErdosRenyi}{2}{True}{False}{False}{all}{Grover}}{0.3}
    &
    \imshow{\datfilename{ErdosRenyi}{4}{True}{False}{False}{all}{Grover}}{0.3}
     & 
    \imshow{\datfilename{ErdosRenyi}{8}{True}{False}{False}{all}{Grover}}{0.3}
    \\
    \bottomrule 
    \end{tabular}
    \caption{Landscape of approximation ratios for depth $p=1$ for the case when k is a power of two.}
    \label{table:landscapesP2}
\end{table}
\endgroup

\begingroup
\setlength{\tabcolsep}{0pt}
\begin{table}[ht]\sffamily
    \centering
    \begin{tabular}{@{\hspace{2pt}}ll@{\hspace{-10pt}}c@{\hspace{-10pt}}c@{\hspace{-10pt}}c@{\hspace{-10pt}}c}
    \toprule
     mixer\phantom{ab}  & $\ket{\phi_0}$\phantom{ab} & $k=3$ & $k=5$ & $k=6$  & $k=7$ \\ 
    \midrule
    X &
    $<k$ &
    \imshow{\datfilename{BarabasiAlbert}{3}{True}{False}{False}{LessThanK}{X}}{0.3}
    &
    \imshow{\datfilename{BarabasiAlbert}{5}{True}{False}{False}{LessThanK}{X}}{0.3}
    &
    \imshow{\datfilename{BarabasiAlbert}{6}{True}{False}{False}{LessThanK}{X}}{0.3}
    &
    \imshow{\datfilename{BarabasiAlbert}{7}{True}{False}{False}{LessThanK}{X}}{0.3}
    \\
    X &
    bal &
    &
    \imshow{\datfilename{BarabasiAlbert}{5}{True}{False}{False}{max_balanced}{X}}{0.3}
    &
    \imshow{\datfilename{BarabasiAlbert}{6}{True}{False}{False}{max_balanced}{X}}{0.3}
    &
    \\
    \midrule
    LX &
    $<k$ &
    \imshow{\datfilename{BarabasiAlbert}{3}{True}{True}{False}{LessThanK}{LX}}{0.3}
    &
    \imshow{\datfilename{BarabasiAlbert}{5}{True}{True}{False}{LessThanK}{LX}}{0.3}
    &
    \imshow{\datfilename{BarabasiAlbert}{6}{True}{True}{False}{LessThanK}{LX}}{0.3}
    &
    \imshow{\datfilename{BarabasiAlbert}{7}{True}{True}{False}{LessThanK}{LX}}{0.3}
    \\
    Grover$^\Box$ &
    $<k$ &
    \imshow{\datfilename{BarabasiAlbert}{3}{True}{True}{False}{LessThanK}{Grovertensorized}}{0.3}
    &
    \imshow{\datfilename{BarabasiAlbert}{5}{True}{True}{False}{LessThanK}{Grovertensorized}}{0.3}
    &
    \imshow{\datfilename{BarabasiAlbert}{6}{True}{True}{False}{LessThanK}{Grovertensorized}}{0.3}
    &
    \imshow{\datfilename{BarabasiAlbert}{7}{True}{True}{False}{LessThanK}{Grovertensorized}}{0.3}
    \\
    Grover &
    $<k$ &
    \imshow{\datfilename{BarabasiAlbert}{3}{True}{True}{False}{LessThanK}{Grover}}{0.3}
    &
    \imshow{\datfilename{BarabasiAlbert}{5}{True}{True}{False}{LessThanK}{Grover}}{0.3}
    &
    \imshow{\datfilename{BarabasiAlbert}{6}{True}{True}{False}{LessThanK}{Grover}}{0.3}
    &
    \imshow{\datfilename{BarabasiAlbert}{7}{True}{True}{False}{LessThanK}{Grover}}{0.3}
    \\
    \bottomrule 
    \end{tabular}
    \caption{Landscape of approximation ratios for depth $p=1$ for the weighted Barabási-Albert graph.}
    \label{tab:landscapeBara}
\end{table}
\endgroup

\begingroup
\setlength{\tabcolsep}{0pt}
\begin{table}[ht]\sffamily
    \centering
    \begin{tabular}{ll@{\hspace{-10pt}}c@{\hspace{-10pt}}c@{\hspace{-10pt}}c@{\hspace{-10pt}}c}
    \toprule
     mixer\phantom{ab}  & $\ket{\phi_0}$\phantom{ab} & $k=3$ & $k=5$ & $k=6$  & $k=7$ \\ 
    \midrule
    X &
    $<k$ &
    \imshow{\datfilename{ErdosRenyi}{3}{True}{False}{False}{LessThanK}{X}}{0.3}
    &
    \imshow{\datfilename{ErdosRenyi}{5}{True}{False}{False}{LessThanK}{X}}{0.3}
    &
    \imshow{\datfilename{ErdosRenyi}{6}{True}{False}{False}{LessThanK}{X}}{0.3}
    &
    \imshow{\datfilename{ErdosRenyi}{7}{True}{False}{False}{LessThanK}{X}}{0.3}
    \\
    X &
    bal &
    &
    \imshow{\datfilename{ErdosRenyi}{5}{True}{False}{False}{max_balanced}{X}}{0.3}
    &
    \imshow{\datfilename{ErdosRenyi}{6}{True}{False}{False}{max_balanced}{X}}{0.3}
    &
    \\
    \midrule
    LX &
    $<k$ &
    \imshow{\datfilename{ErdosRenyi}{3}{True}{True}{False}{LessThanK}{LX}}{0.3}
    &
    \imshow{\datfilename{ErdosRenyi}{5}{True}{True}{False}{LessThanK}{LX}}{0.3}
    &
    \imshow{\datfilename{ErdosRenyi}{6}{True}{True}{False}{LessThanK}{LX}}{0.3}
    &
    \imshow{\datfilename{ErdosRenyi}{7}{True}{True}{False}{LessThanK}{LX}}{0.3}
    \\
    Grover$^\Box$ &
    $<k$ &
    \imshow{\datfilename{ErdosRenyi}{3}{True}{True}{False}{LessThanK}{Grovertensorized}}{0.3}
    &
    \imshow{\datfilename{ErdosRenyi}{5}{True}{True}{False}{LessThanK}{Grovertensorized}}{0.3}
    &
    \imshow{\datfilename{ErdosRenyi}{6}{True}{True}{False}{LessThanK}{Grovertensorized}}{0.3}
    &
    \imshow{\datfilename{ErdosRenyi}{7}{True}{True}{False}{LessThanK}{Grovertensorized}}{0.3}
    \\
    Grover &
    $<k$ &
    \imshow{\datfilename{ErdosRenyi}{3}{True}{True}{False}{LessThanK}{Grover}}{0.3}
    &
    \imshow{\datfilename{ErdosRenyi}{5}{True}{True}{False}{LessThanK}{Grover}}{0.3}
    &
    \imshow{\datfilename{ErdosRenyi}{6}{True}{True}{False}{LessThanK}{Grover}}{0.3}
    &
    \imshow{\datfilename{ErdosRenyi}{7}{True}{True}{False}{LessThanK}{Grover}}{0.3}
    \\
    \bottomrule 
    \end{tabular}
    \caption{Landscape of approximation ratios for depth $p=1$ for the unweighted Erdős-Rényi graph.}
    \label{tab:landscapeErdo}
\end{table}
\endgroup

\end{appendix}

\end{document}